%% file: main.tex
\documentclass{article}
\usepackage{spconf,amsmath,graphicx}

\usepackage[utf8]{inputenc}
\usepackage[T1]{fontenc}  
\usepackage{kotex}

\usepackage[numbers,sort]{natbib}  

\usepackage{url}
\usepackage{cleveref}
\Crefname{figure}{Fig.}{Figs.}
\Crefname{subsection}{Subsection}{Subsections}

\usepackage{indentfirst}
\usepackage[skip=5pt, indent=10pt]{parskip}

\usepackage{caption}
\usepackage{subcaption}
\usepackage{tabularx}
\usepackage{multirow}
\usepackage{diagbox}
\newcolumntype{L}[1]{>{\raggedright\let\newline\\\arraybackslash\hspace{0pt}}m{#1}}
\newcolumntype{C}[1]{>{\centering\let\newline\\\arraybackslash\hspace{0pt}}m{#1}}
\newcolumntype{R}[1]{>{\raggedleft\let\newline\\\arraybackslash\hspace{0pt}}m{#1}}

\usepackage{graphicx}
\usepackage{wrapfig}
\usepackage[export]{adjustbox}

\usepackage{xcolor}  

\title{OLKAVS: An Open Large-Scale Korean Audio-Visual Speech Dataset}
\name{
    \begin{tabular}{@{}c@{}}
    Jeongkyun Park$^1$ \qquad 
    Jung-Wook Hwang$^2$ \qquad 
    Kwanghee Choi$^3$ \qquad 
    Seung-Hyeon Lee$^2$\\ 
    Jun Hwan Ahn$^4$ \qquad 
    Rae-Hong Park$^{2,5}$ \qquad 
    Hyung-Min Park$^{1,2,5,\dag}$
    \end{tabular}
    \thanks{
    \dag~ Corresponding author: \texttt{hpark@sogang.ac.kr}. \\
    This work was partly supported by National Information Society Agency (NIA) grant (AI HUB Voice and Motion Data - Lip Voice) and Institute of Information \& communications Technology Planning \& evaluation (IITP) grant funded by the Korea government (MSIT) (No. 2022-0-00989, Development of Artificial Intelligence Technology for Multi-speaker Dialog Modeling). The audio-visual dataset collection project was proposed based on the National Information Society Agency (NIA)'s request for proposal (AI HUB Voice and Motion Data - Lip Voice), funded by the Korea government (MSIT). After obtaining consent from the participants, the dataset was collected and published on https://aihub.or.kr, an official portal that discloses various AI data. All of this was done under the supervision of the NIA.}
}
\address{$^1$ Department of Artificial Intelligence, Sogang University, Seoul 04107, Republic of Korea\\
         $^2$ Department of Electronic Engineering, Sogang University, Seoul 04107, Republic of Korea \\
         $^3$ Language Technologies Institute, Carnegie Mellon University, Pittsburgh, PA 15213, USA \\
         $^4$ Mindslab Inc., Gyeonggi-do 13493, Republic of Korea \\
         $^5$ ICT Convergence Disaster/Safety Research Institute, Sogang University, Seoul 04107, Republic of Korea \\}
\begin{document}
%
\maketitle
%

\input{sections/0_abstract}
\input{sections/1_intro}
\input{sections/3_dataset}
\input{sections/4_exp}
\input{sections/5_concl}

\clearpage
\section{REFERENCES}
\label{sec:refs}

\renewcommand*{\bibfont}{\small}
\setlength{\bibsep}{0.0pt}
\bibliographystyle{IEEEbib}
\bibliography{refs}

\newpage
\appendix

\input{appendix/distribution}

\end{document}

%% file: sections/0_abstract.tex
\begin{abstract}
Inspired by humans comprehending speech in a multi-modal manner, various audio-visual datasets have been constructed.
However, most existing datasets focus on English, developed from pre-existing videos using various prediction models,
and have only a small number of multi-view videos.
To mitigate the limitations, we constructed the Open Large-scale Korean Audio-Visual Speech (OLKAVS) dataset, which is the largest among publicly available audio-visual speech datasets.
The dataset contains 1,150 hours of transcribed audio from 1,107 Korean speakers in a studio setup with nine different viewpoints and various noise situations.
We also provide the pre-trained baseline models for two tasks: audio-visual speech recognition and lip reading.
We conducted experiments based on the models to verify the effectiveness of multi-modal and multi-view training over uni-modal and frontal-view-only training.
We expect the OLKAVS dataset to facilitate multi-modal research in broader areas.
\end{abstract}

\vspace{-0.5em}
\begin{keywords}Audio-visual speech datasets, multi-view datasets, lip reading, audio-visual speech recognition, deep learning
\end{keywords}
\vspace{-1em}

%% file: sections/1_intro.tex
\vspace{-0.5em}
\section{Introduction}\label{sec:intro}
\vspace{-0.5em}

\input{src/table/databases.tex}

People comprehend speech in a multi-modal manner in real-world communication, not only relying on listening to utterances but also reading speakers' faces and lips to improve perception.
Inspired by this, many audio-visual speech datasets have been introduced, capturing both utterances and talking faces in a synchronized manner.
These datasets enabled the research on various speech-related applications, such as noise-robust speech recognition \cite{ma2021end}, 
lip reading \cite{chung2016lip,chung2017lip}, mouth motion analysis \cite{anina2015ouluvs2}, and speaker recognition \cite{chung2018voxceleb2}.
To further embolden future research, it is essential to develop datasets with more diverse situations, such as various speakers \cite{yang2019lrw,chung2017lip,afouras2018lrs3}, languages \cite{yang2019lrw,schwiebert-etal-2022-multimodal,chung2018voxceleb2}, sentences \cite{chung2017lip,afouras2018lrs3}, and multiple viewpoints of the head \cite{anina2015ouluvs2,chung2017profile}.

\vspace{-0.2em}
For the research on lip reading, multiple datasets \cite{matthews2002extraction, patterson2002cuave, zhao2009lipreading, cooke2006audio} were constructed, albeit the limited vocabulary and sample size of classical datasets, containing less than 50 speakers.
OuluVS2 \cite{anina2015ouluvs2} and AVICAR \cite{lee2004avicar} considered full sentences with more various vocabularies, more than a thousand words.
Interestingly, they also provided multi-view videos from four and five different angles, respectively.

\vspace{-0.2em}
To develop larger datasets for word utterances and reflect real-world speech with unrestricted vocabulary, LRW proposed a data collection pipeline to label TV news footage automatically \cite{chung2016lip}.
The pipeline-based approach has influenced numerous follow-up sentence-based audio-visual speech datasets: LRS2-BBC \cite{chung2017lip,afouras2018deep}, LRS3-TED \cite{afouras2018lrs3}, and MV-LRS \cite{chung2017profile}.
MV-LRS primarily focused on providing multi-view video, dividing the data into five categories based on the face angle to facilitate the multi-view evaluation.
Leveraging the development of the collection pipeline from \cite{chung2016lip}, some non-English datasets were introduced such as LRW-1000 \cite{yang2019lrw}, GLips \cite{schwiebert-etal-2022-multimodal}, and MISP2021 \cite{chen2022audio}.
VoxCeleb2 \cite{chung2018voxceleb2} also contains multilingual speech in more than 100 languages, but there are no transcriptions available in the dataset.

\vspace{-0.2em}
Despite the development in audio-visual datasets, only a few attempts have been made to collect Korean speech.
\cite{jo2000collection} collected predefined syllables of a single speaker with 7 views.
Also, \cite{lee2008robust} and \cite{lee2019visual} collected predefined word utterances, such as digits and city names, of 56 and 9 speakers, respectively.
Unfortunately, the size of all the datasets is too minuscule to support deep learning-driven models; moreover, some are not publicly available.

Despite the increasing attention to the audio-visual speech domain, existing datasets have a number of limitations. \linebreak
1) Foremost, most datasets, especially large-scale datasets, only focus on English alone, with a few exceptions.
2) Even though many large-scale datasets have been introduced, they often rely on publicly available videos on the TV and web, introducing an inherent dependency on various detection models in collection pipelines and sometimes encountering copyright restrictions.
3) Many speech datasets with multi-view videos are too small for modern deep learning, with less than 100 hours.
The MV-LRS \cite{chung2017profile} stands out with its 777 hours of face rotations, but it offers only one view at a time.
Refer to \Cref{table:database} for more details.

To mitigate the limitations, we developed a new audio-visual speech dataset, \textbf{Open Large-scale Korean Audio-Visual Speech (OLKAVS)} dataset.
We gathered 1,150 hours of audio from 1,107 speakers in a studio setup with corresponding Korean transcriptions.
We recorded a speaker from 9 viewpoints: frontal, cross (left, right, up, and down), and diagonal (upper left, upper right, lower left, and lower right).
The OLKAVS' advantages can be summarized as follows.
\begin{enumerate}
  \item It is the only Korean audio-visual speech dataset with a significant amount of more than hundreds of hours while many datasets focus on English.
  \vspace{-0.5em}
  \item Among publicly available datasets, it is the largest audio-visual speech dataset. In particular, it provides complete manual transcriptions for all utterances.
  \vspace{-0.5em}
  \item It is a large-scale dataset providing synchronously recorded multi-view videos.
\end{enumerate}
OLKAVS can advance research in multi-modal tasks like audio-visual speech recognition (AVSR) and lip reading, as well as in speech-only tasks such as speech and speaker recognition.
For AVSR and lip-reading, we have publicly released the pre-trained baseline models to underscore the dataset's value and assist future research.\footnote{The dataset and the pre-trained models are available at \texttt{\url{https://aihub.or.kr/aihubdata/data/view.do?currMenu=115&topMenu=100&aihubDataSe=realm&dataSetSn=538}} and \texttt{\url{https://github.com/IIP-Sogang/olkavs-avspeech}}, respectively.}

In this paper, we aim to provide a complete guideline for the dataset.
The details of how the dataset was constructed are enumerated in \Cref{sec:dataset}.
Also, we demonstrated our dataset works on two downstream tasks, AVSR and multi-view lip reading in \Cref{sec:exp}.

%% file: src/table/databases.tex
\begin{table*}[!h]
\newcommand\view{1.5}
\newcommand\cont{4.7}
\centering

\caption{
Summarized statistics of various audio-visual speech datasets.
The largest values are written in bold.
Some datasets containing various head poses without explicitly fixed views are listed under `unfixed'.
Hours, \#Utt., and \#Subj. of LRS2-BBC and LRS3-TED are approximations that can be slightly larger due to potential overlaps in the dataset.
}
\begin{tabular}{ C{2.5cm} R{2cm} R{1cm} R{1cm} C{\view cm} C{2cm} L{\cont cm} }
 \hline
 Database & Hours & \#Utt. & \#Subj. & Views & Language & Content \\ 
 \hline
 OuluVS2~\cite{anina2015ouluvs2}
    & - & 1.5K & 53 &
    \begin{tabular} {@{}C{\view cm}@{}}
        5 views
    \end{tabular} & English & 
    \begin{tabular} {@{}L{\cont cm}@{}}
        digits, phrases, 
        TIMIT sentences 
    \end{tabular} \\
 \hline
 AVICAR~\cite{lee2004avicar}
    & - & 59K & 86 &
    \begin{tabular} {@{}C{\view cm}@{}}
        4 views
    \end{tabular} & English & 
    \begin{tabular} {@{}L{\cont cm}@{}}
        digits, letters,  
        TIMIT sentences
    \end{tabular} \\
 \hline
 LRS2-BBC~\cite{afouras2018deep}
    & 224.5 & 144K & - & unfixed & English & sentences \\
 \hline
 LRS3-TED~\cite{afouras2018lrs3}
    & 438 & 152K & \textbf{9,545} & unfixed & English & sentences \\
 \hline
 MV-LRS~\cite{chung2017profile}
    & 777.2 & 504K & - & 
    \begin{tabular} {@{}C{\view cm}@{}}
        5 views 
    \end{tabular} & English & sentences \\
 \hline
 VoxCeleb2~\cite{chung2018voxceleb2}
    & 2,442 & 1.1M & 6,112 & unfixed & Multilingual & speaker info. only \\
 \hline
 GLips
 ~\cite{schwiebert-etal-2022-multimodal} 
    & 81 & 250K & - & unfixed & German & words \\
 \hline
 MISP2021~\cite{chen2022audio}          
    & 141.2 & - & 263 & 
    \begin{tabular} {@{}C{\view cm}@{}}
        3 views
    \end{tabular} &  Mandarin  & sentences \\
 \hline
 Jo \textit{et al.} \cite{jo2000collection}
    & - & 168 & 1 & 
    \begin{tabular} {@{}C{\view cm}@{}}
        7 views
    \end{tabular} & Korean & syllables, consonants, vowels \\
 \hline
 Lee \textit{et al.} \cite{lee2008robust}
    & - & 4.5K & 56 & frontal & Korean & digits, words \\
 \hline
 OLKAVS & 
    \begin{tabular} {@{}R{2 cm}@{}}
         Video: \bf{5,750} \\
        Audio: 1,150 \\
    \end{tabular} & \bf{2.5M} & 1,107 & 
    \begin{tabular} {@{}C{\view cm}@{}}
        \bf{9 views} 
    \end{tabular} & Korean & 
    \begin{tabular} {@{}L{\cont cm}@{}}
        sentences
    \end{tabular} \\
 \hline
\end{tabular}
\vspace{-1.5em}
\label{table:database}
\end{table*}

%% file: sections/3_dataset.tex
\vspace{-0.5em}
\section{OLKAVS Dataset}\label{sec:dataset}
\vspace{-0.5em}

In this section, we describe our dataset in detail.
OLKAVS contains 1,150 and 5,750 hours of audio and video in total, respectively, where we recorded five videos per single audio.
To the best of our knowledge, this is the largest audio-visual speech dataset, especially for multi-view datasets.

\vspace{-1em}
\subsection{Scripts and Speakers}\label{subsec:script_n_speaker}

The OLKAVS dataset contains 14 different topics, where we constructed a set of sentences based on their corresponding keywords.
Each topic has at least 2,000 sentences, so the total sentences exceed 28,000.
Each sentence is written in the Korean language.
Then, randomly chosen sentences from each topic are combined to build a script with an average of 2,750 characters (5 minutes long on average).
Generated scripts are provided to the speakers when recording.

A total of 1,107 speakers participated in constructing the OLKAVS dataset, and all the speakers formally consented to using portrait rights and disclosing personal information.
They were recruited while considering the demography of biological sex and age.
The numbers of males and females are balanced with 555 males and 552 females. 
The age of speakers ranges from 10 to 60 and above, following its distribution within Korea.
Note that the teenagers were specifically asked to speak on six topics chosen from the total set of 14, as these topics were deemed more understandable for younger people (e.g., food, education/school), while the others spoke on all 14 topics.

30\% of the speakers are speech experts, such as announcers, actors, would-be announcers, and would-be actors.
Also, 50\% of the experts' speeches were done spontaneously about the provided news article, while it was not collected from the other speakers due to the difficulty of spontaneous speech.

\input{src/fig/view_sample.tex}

\subsection{Recording Environment}\label{subsec:recording}

Each utterance was recorded from five different views simultaneously by GoPro Hero7 cameras.
As shown in \Cref{fig:viewsample}, the captured views comprise two types of recorded views: cross views (frontal, up, down, left, and right) and diagonal views (frontal, upper left, upper right, lower left, and lower right).
Notably, the frontal view is present in both categories.
Unlike prior datasets \cite{anina2015ouluvs2,chung2017profile,jo2000collection} that only consider horizontal views, i.e., left, frontal, and right, OLKAVS dataset also considers the vertical variants, i.e., upper left, upper center, upper right, lower left, lower center, and lower right.

A condenser microphone was employed for recording speech in stereo, ensuring a consistent distance of about 1 m between the speaker and microphone, as well as the frontal-view camera.
A soundproof booth measuring about 1.5 m $\times$ 1.5 m $\times$ 2.0 m was utilized to capture the speech.
To simulate noisy conditions inside the studio environment, we played the pre-recorded noise inside the studio.
We constructed six noise scenarios: no noise, indoor ambiance, indoor noise (laughing and clapping), traffic noise, construction site noise, and natural outdoor noise (flowing water and animal sounds).
Noise-free scenarios account for 29\% of total hours, each of the others making up 14\%.
All the noise audios were collected from the corresponding real-world situations.

\subsection{Data Cleansing and Post-processing}\label{subsec:postprocess}
After recording, we transcribed all the scripted and spontaneous speech.
Even though the scripts were given to the speakers, some of the speakers mispronounced a few words or spoke pause fillers such as `um' or `ah,' where everything was transcribed afterward.
All transcriptions are the ground truth for speech, being provided only in Korean language.

We manually filtered out the problematic recordings by the following failure criteria: the entire lip or facial region is not recorded, the filming environment is too dark, the noise level is too high, or videos and the audio are not in sync.
We also trimmed every audio-visual pair to have the same length.

The OLKAVS dataset provides predefined splits of train, validation, and evaluation with a ratio of 8:1:1.
The dataset is split based on speaker ID so that no speaker co-occurs on different splits.
Only the train and validation sets are made publicly available.
For ease of use, we also provide the bounding box coordinates of the lip and the face.
We used the face alignment estimation module from \cite{bulat2017far}.
Facial landmarks were extracted every 3 seconds and linearly interpolated for the frames in between.

Videos are stored in MP4 format with the FHD (1,920$\times$1,080) resolution with 30 fps, while audios are encoded in WAV format with stereo channels and 48 kHz sampling rates.
Label data include transcriptions, the utterances' start and end times, speaker information on ID, age, gender, and whether the speaker is an expert.

%% file: src/fig/view_sample.tex
\begin{figure}
    \centering
    \begin{subfigure}[b]{0.45\textwidth}
         \centering
         \includegraphics[width=0.9\columnwidth]{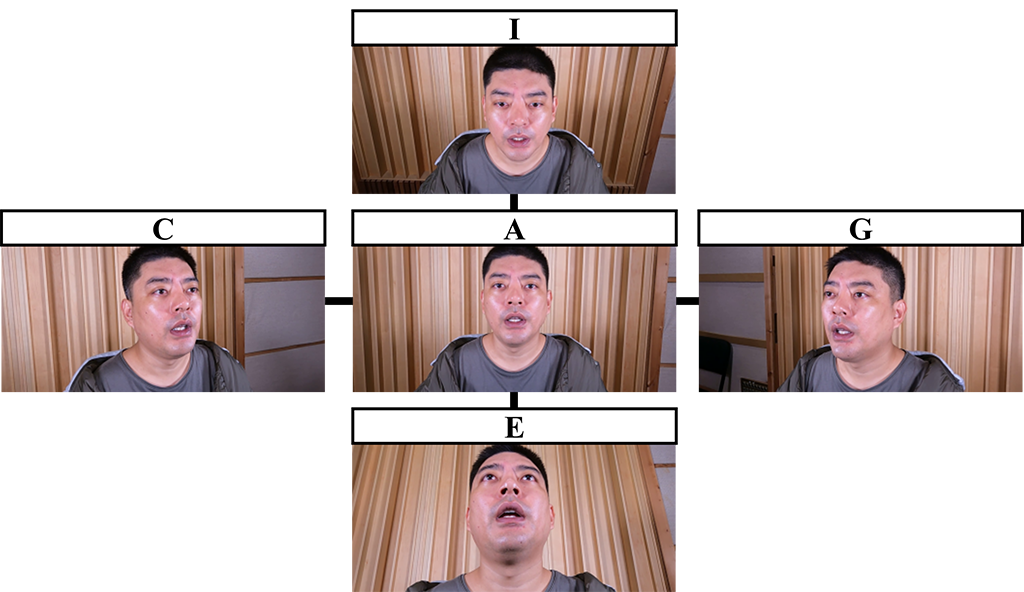}
         \caption{Cross views}
         \label{fig:view_ppd}
    \end{subfigure}
    \hfill
    \begin{subfigure}[b]{0.45\textwidth}
         \centering
         \includegraphics[width=0.9\columnwidth]{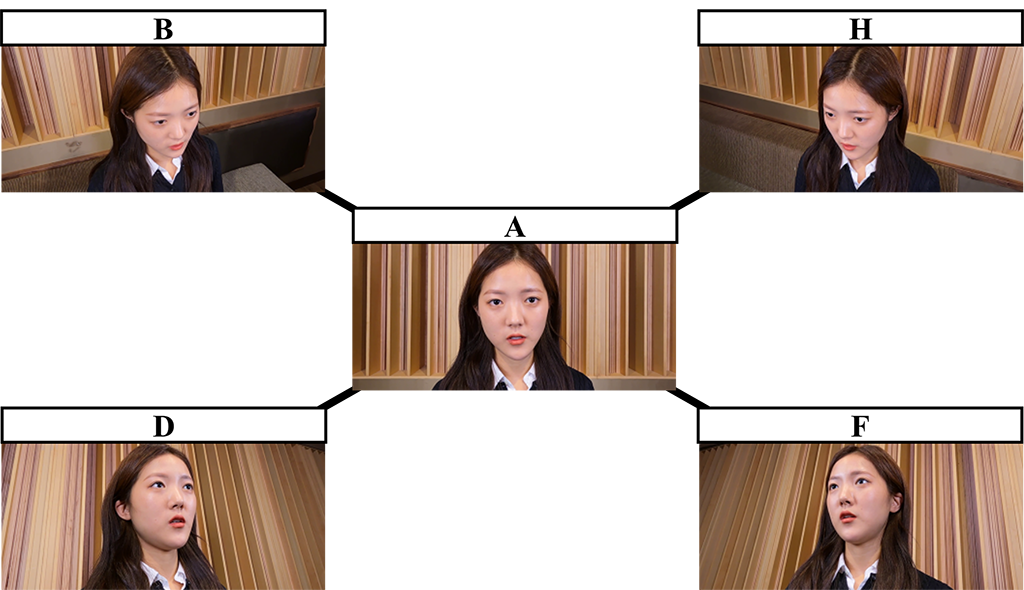}
         \caption{Diagonal views}
         \label{fig:view_dgn}
    \end{subfigure}
    \vspace{-0.5em}
    \caption{
Video samples from the OLKAVS dataset.
Five different views were synchronously recorded for each recording session.
(a) and (b) show two types of views and denote their labels.
}
    \label{fig:viewsample}
\vspace{-1.5em}
\end{figure}

%% file: sections/4_exp.tex
\input{src/table/exp1_baseline.tex}
\input{src/table/exp2_multiview.tex}

\vspace{-0.5em}
\section{Experiments}\label{sec:exp}
\vspace{-0.5em}
In this section, we outline OLKAVS dataset baselines for AVSR and multi-view lip reading. 
The former examines multi-modal versus uni-modal training, while the latter assesses the advantage of multi-view data over frontal-view-only training.

In both experiments, videos were cropped with the lip-bounding boxes we provided.
Then, we split the speech sentence-wise, based on the timestamps in the transcripts.
We tokenized the transcripts into grapheme units of 21 vowels, 19 initial consonants, and 27 final consonants, removed punctuations, and converted numbers into corresponding words.

\vspace{-1em}
\subsection{AVSR}\label{subsec:speech_recog}

We employed the conformer sequence-to-sequence (CM-Seq2seq) architecture \cite{ma2021end} for the audio-visual model (AV-model), using 8-head attention and 256 feature dimensions for the 12-layered conformer encoder and 6-layered transformer decoder.
For visual processing, we utilized a pre-trained front-end from \cite{afouras2018deep} trained on several English lip reading datasets \cite{chung2016lip,chung2017profile,afouras2018deep} and froze them during training.

During training, we introduced additional noise from DEMAND \cite{thiemann2013demand} and NOISEX-92 \cite{varga1993assessment} to simulate more challenging environments than those encountered in our recordings, including scenarios with superimposed noise.
They varied six different input signal-to-noise ratios (SNRs): $-$5 dB, 0 dB, 5 dB, 10 dB, 15 dB, and 20 dB.
We used the Adam optimizer \cite{kingma15adam} with a 0.0001 learning rate and the Noam learning rate scheduler \cite{vaswani2017transformer}.
Furthermore, for the audio-only (A-model) and visual-only (V-model) baselines, we simply removed the corresponding encoder and the fusion module, following \cite{ma2021end}.

For the baseline evaluation (\Cref{table:baseline}), we evaluated all three models, A-model, V-model, and AV-model, on the publicly available validation set containing various noises, as mentioned in \Cref{subsec:recording}.
To evaluate the noise robustness of the models, we varied the noise levels explicitly by mixing the noise-free validation subset with babble noise from NOISEX-92 \cite{varga1993assessment}.
The performance was measured in terms of character error rate (CER), word error rate (WER), and space-normalized word error rate (sWER) 
\cite{bang2020ksponspeech},
which were computed for both Korean characters and words.

We provided speech recognition baselines for the OLKAVS dataset in \Cref{table:baseline}, which contains the first published outcomes for the Korean AVSR and lip reading at the sentence level.
The AV-model demonstrated a notable performance improvement over the A-model under noisy conditions while exhibiting comparable performance with the A-model for the clean data, indicating the effectiveness of utilizing audio-visual modalities in speech recognition tasks.

\subsection{Multi-view Lip Reading}\label{subsec:lip_read}
We adopted the V-model architecture and the optimization settings from \Cref{subsec:speech_recog} but trained it from scratch to thoroughly examine the visual representations.

To assess the benefit of using all the views in enhancing the lip-reading performance \cite{lee2017multi}, 
we compared how the lip-reading models of the same architecture perform when trained only with the frontal view (F-model) versus using all the views (All-model).
For evaluation, we used a noise-free subset of the validation data.

\Cref{table:multiview} demonstrates that the frontal lip reading performance of the All-model (32.16\%) was superior to that of the F-model (41.24\%) with an 22\% relative reduction in CER, which reaffirmed the results of \cite{lee2017multi} that utilizing multi-view training data was superior to that of the frontal-view-only.
We also analyzed the difficulty of lip reading from each view by comparing the performance on various views.
Confirming \cite{lee2017multi}, centered views underperformed right and left views in lip reading, except at the Medium-high level. 
This may result from model bias towards the frontal view, which has double the training data of the others.
It is noteworthy that its performance was better than that of the upper and medium-high angles since the lower angle showed lip movement better.
These results underscore the promise of using non-frontal views, guiding our future works to leverage all synchronized views for enhanced lip representations.

%% file: src/table/exp1_baseline.tex
\begin{table*}[t]
\centering
\caption{
Baseline performances of A-model, V-model, and AV-model on the OLKAVS validation set. `Val.' refers to the full validation dataset published on the official portal. `Clean' indicates a noise-free validation subset while `(number) dB' represents the same subset, but corrupted by adding babble noise from NOISEX-92 \cite{varga1993assessment} to get the designated input SNR
}
\small
\setlength{\tabcolsep}{0pt} 
\renewcommand{\arraystretch}{1.5}
\makebox[\linewidth][c]{
\resizebox{\linewidth}{!}{
\begin{tabularx}{\paperwidth}{
    *{1}{>{\centering\arraybackslash}X||}*{6}{>{\centering\arraybackslash}X}*{1}{>{\centering\arraybackslash}X|}*{6}{>{\centering\arraybackslash}X}*{1}{>{\centering\arraybackslash}X|}*{6}{>{\centering\arraybackslash}X}*{1}{>{\centering\arraybackslash}X|}
}
    \hline
    \multirow{2}{*}{Model} & \multicolumn{7}{c|}{CER (\%)} & \multicolumn{7}{c|}{WER (\%)} & \multicolumn{7}{c|}{sWER (\%)} \\
    \cline{2-22}
       & Val. & Clean & 15dB & 10dB & 5dB & 0dB & $-5$dB 
       & Val. & Clean & 15dB & 10dB & 5dB & 0dB & $-5$dB 
       & Val. & Clean & 15dB & 10dB & 5dB & 0dB & $-5$dB  \\
    \hline
    A  & ~\textbf{3.57} & ~2.00 & ~2.29 & ~2.89 & ~4.79 & 11.15 & 30.76 
       & \textbf{10.61} & \textbf{7.27} & 7.96 & 9.24 & 12.97 & 24.33 & 54.03
       & \textbf{8.11} & \textbf{4.36} & 5.13 & 6.47 & 10.73 & 23.82 & 58.72 \\
    V  & 26.64 & 25.08 & 25.08 & 25.08 & 25.08 & 25.08 & 25.08 
       & 47.89 & 45.20 & 45.20 & 45.20 & 45.20 & 45.20 & 45.20
       & 50.00 & 46.41 & 46.41 & 46.41 & 46.41 & 46.41 & 46.41 \\
    AV & ~3.64 & ~\textbf{1.98} & ~\textbf{2.24} & ~\textbf{2.70} & ~\textbf{4.09} & ~\textbf{8.42} & \textbf{19.75} 
       & 10.82 & ~7.29 & ~\textbf{7.83} & ~\textbf{8.88} & \textbf{11.87} & \textbf{20.34} & \textbf{39.97}
       & ~8.18 & ~4.37 & ~\textbf{5.01} & ~\textbf{6.08} & ~\textbf{9.32} & \textbf{18.54} & \textbf{39.57} \\
    \hline
\end{tabularx}
}
}
\label{table:baseline}
\end{table*}

%% file: src/table/exp2_multiview.tex
\begin{table*}[!t]
\centering
\caption{CERs (\%) of lip reading on a subset of validation data with no noise in the OLKAVS dataset}
\begin{tabular}{|l|c c c|c c c|c c c|}
 \hline
 \multirow{2}{*}{\backslashbox[35mm]{Model}{View}}    & \multicolumn{3}{|c|}{Upper} & \multicolumn{3}{|c|}{Medium-high} & \multicolumn{3}{|c|}{Lower} \\\cline{2-10}
&left&center&right&left&center (frontal)&right&left&center&right\\
 \hline
 \hline
 F-model (Frontal only)  & 73.58 & 75.99 & 77.88 & 63.19 & 41.24 & 71.78 & 59.22 & 
53.99 & 61.99 \\
 All-model (All views)   & \textbf{41.02} & \textbf{46.14} & \textbf{42.43} & \textbf{34.93} & \textbf{32.16} & \textbf{34.60} & \textbf{31.91} & \textbf{33.55} & \textbf{32.21} \\
 \hline
\end{tabular}
\vspace{-1em}
\label{table:multiview}
\end{table*}

%% file: sections/5_concl.tex
\section{Conclusion}\label{sec:concl}
\vspace{-0.5em}
In this paper, we introduced a novel audio-visual speech dataset, the Open Large-scale Korean Audio-Visual Speech (OLKAVS) dataset, which is the largest publicly available audio-visual speech dataset, to the best of our knowledge.
We also provided Korean baselines for AVSR and multi-view lip reading, 
demonstrating that using both multi-modal and multi-view data during training improved upon the uni-modal and frontal-view-only data.
Finally, we emphasize that the OLKAVS dataset can be utilized for broader research areas outside of AVSR and lip reading, such as Korean speech recognition, speaker recognition, and mouth motion analysis.

%% file: appendix/distribution.tex
\onecolumn
\section*{Appendix}
\section{Data distribution}
\input{src/table/dataset-age}

\input{src/table/dataset-topic}

\newpage
\section{Sample scripts}
\input{src/table/script}

%% file: src/table/dataset-age.tex
\begin{table}[h]
\centering
\caption{Age distribution of speakers in the OLKAVS dataset}
\begin{tabular}{ C{2cm} C{2cm} C{2.5cm} }
 \hline
 Age $a$ & \# of speakers & Speaker ratio (\%) \\
 \hline\hline
 $10 \leq a < 20$     & 109 & 10 \\
 \hline
 $20 \leq a < 30$     & 286 & 26 \\
 \hline
 $30 \leq a < 40$     & 280 & 25 \\
 \hline
 $40 \leq a < 50$     & 259 & 23 \\
 \hline
 $50 \leq a < 60$     & ~94  & ~8 \\
 \hline
 $a \geq 60$  & ~79  & ~7 \\
 \hline
\end{tabular}
\label{table:age}
\vspace{-1em}
\end{table}

%% file: src/table/dataset-topic.tex
\begin{table*}[!ht]
\centering
\caption{
Script topics of the OLKAVS dataset.
Only the topics marked with an asterisk (*) were presented to the teenagers, as other topics might be challenging to understand fully
}
\begin{tabular}{ C{2.2cm} C{1cm} C{1.3cm} C{10cm} }
 \hline
 Topic & Hours & Ratio (\%) & Keyword examples \\
 \hline\hline
 Health/Diet* 
 & 497 & 8.7 & Disease, Hospital, Medicine, Healthcare, Side Effects, Workout, Treatment \\
 \hline
 Daily 
 & 484 & 8.5 & Weather, Hobby, Habit, Self-management, Shopping, Leisure, Club, Recollection \\
 \hline
 Job/Work
 & 480 & 8.4 & Recruiting, Job, Company, Industries, Employment, Labor, Foundation \\
 \hline
 Financial
 & 445 & 7.7 & Price, Cost, Tax, Investment, Stock, Real Estate, Deposit, Saving, Loan, Fund, Interest Rate, Exchange Rate \\
 \hline
 Human Relations
 & 413 & 7.2 & Family, Friends, Couple, Quarrel, Reconciliation \\
 \hline
 Pet
 & 409 & 7.1 & Dog, Cat, Adoption, Pet Disease, Memories with Pet, Pet Accessory \\
 \hline
 Sports/Leisure*
 & 406 & 7.1 & Player, Watching a Game, Olympic, Paralympic, National Team, Rule, Water Sports, Climbing \\
 \hline
 Food*
 & 399 & 6.9 & Restaurants, Ingredients, Recipe, Local Foods, Fruits, Vegetables, Fishes, Korean Foods, World Foods, Delivery, Meal Kit\\
 \hline
 Education/School*
 & 394 & 6.8 & Examination, Subjects, Major, Quiz, Grades, Entrance, Graduation, School, Teacher, Professor, Mid Term, Club \\
 \hline
 Vacation
 & 384 & 6.7 & Trip, Travel, Destination, Accommodation, Photo, Holiday, Beach, Attraction, Christmas\\
 \hline
 Art/Culture*
 & 388 & 6.7 & Book, Music, Broadcast, Movie, Drama, Celeb, Singer, Actor, Show, Museum, Musical, Poem, Writer, Exhibition \\
 \hline
 Transportation*
 & 379 & 6.6 & Driving, Car, Bus, Taxi, Subway, Train, Airplane, Bicycle, Scooter, Ticket Reservation, Public Transit \\
 \hline
 Social Issues
 & 357 & 6.2 & Social Issue, News, Human Right, Journalism, Global Warming, Environment, Animal Welfare \\
 \hline
 \begin{tabular}{@{}c@{}}
    Ceremonial \\ Occasions
 \end{tabular}
 & 307 & 5.3 & Wedding, Invitation, Visitor, Presents, Funeral, Condolences, Rite\\
 \hline
\end{tabular}
 \label{table:topics}
\end{table*}

%% file: src/table/script.tex
\begin{table*}[!ht]
\centering
{
\scriptsize

\caption{
A sample sentence per topic and a sample guideline given to speakers for the OLKAVS dataset.
Only the expert speakers received guidelines for spontaneous speech, whereas scripts were given to the average speakers (Refer to \Cref{subsec:script_n_speaker} for more details.).
English translations are provided in the table for ease of understanding, but the speakers were only supplied with Korean sentences and guidelines
}
\label{table:script}

\resizebox{\textwidth}{!}{
\begin{tabular}{ |C{2.2cm}||C{3cm}|p{10.3cm}| }
 \hline
 \multirow{15}{*}[-5em]{
    \begin{tabular} {@{}c@{}}
        Script \\
        (for scripted \\
        speech)
    \end{tabular}
 }
    & 주제 (Topic) & \multicolumn{1}{c|}{스크립트 (Script)}
    \\\cline{2-3}
    & \begin{tabular} {@{}c@{}}
        건강/다이어트 \\
        (Health/Diet)
    \end{tabular}
    & \begin{tabular} {@{}l@{}}
        딸이 약을 제대로 챙겨 먹지 않아서 건강 상태가 더 악화됐어. \\
        (My daughter's health condition worsened because she didn't take her medicine properly.)
    \end{tabular}
    \\\cline{2-3}

    & \begin{tabular} {@{}c@{}}일상생활\\(Daily) \end{tabular}
    & \begin{tabular} {@{}l@{}}
        이번에 새로 산 바지가 한 번 빨았는데 줄어들어서 너무 작아졌어요. \\
        (The new pants I bought were washed once, but they shrunked too small.)
    \end{tabular}
    \\\cline{2-3}

    & \begin{tabular} {@{}c@{}}
        일/직장/직업\\
        (Job/Work)
    \end{tabular}
    & \begin{tabular} {@{}l@{}}
        요즘 직장 생활은 위아래로 부딪혀서 업무에 집중하기 힘들어.\\
        (These days, work life is bumping up and down, so it's hard to work properly.)
    \end{tabular}
    \\\cline{2-3}

    & \begin{tabular} {@{}c@{}}
        경제/재테크 \\
        (Financial)
    \end{tabular}
    & \begin{tabular} {@{}l@{}}
        지금 달러가 강세라고 하던데 그럼 내가 뭘 해야지 이득을 볼 수 있는 거야? \\
        (They say the dollar is strong right now, so what should I do to benefit?)
    \end{tabular}
    \\\cline{2-3}

    & \begin{tabular} {@{}c@{}}
        인간관계 \\
        (Human Relations)
    \end{tabular}
    & \begin{tabular} {@{}l@{}}
        남편이 나를 너무 못살게 굴어서 결혼생활이 불행해. \\
        (My husband treats me so badly that my marriage is unhappy.)
    \end{tabular}
    \\\cline{2-3}

    & \begin{tabular} {@{}c@{}}
        반려동물 \\
        (Pet)
    \end{tabular}
    & \begin{tabular} {@{}l@{}}
        동물 병원에서 일하다 보니 나도 반려동물을 키우고 싶어져. \\
        (Working at a veterinary hospital makes me want to have a pet too.)
    \end{tabular}
    \\\cline{2-3}

    & \begin{tabular} {@{}c@{}}
        스포츠/레저 \\
        (Sports/Leisure)
    \end{tabular}
    & \begin{tabular} {@{}l@{}}
        넌 제일 좋아하는 축구 선수가 누구야? \\
        (Who is your favorite soccer player?)
    \end{tabular}
    \\\cline{2-3}

    & \begin{tabular} {@{}c@{}}
        음식 \\
        (Food)
    \end{tabular}
    & \begin{tabular} {@{}l@{}}
        오늘 집에서 스테이크를 미디엄 레어로 구워 먹었어. \\
        (I had a steak cooked medium rare at home today.)
    \end{tabular}
    \\\cline{2-3}

    & \begin{tabular} {@{}c@{}}
        진학/학교 \\
        (Education/School)
    \end{tabular}
    & \begin{tabular} {@{}l@{}}
        나는 벌써 휴학 신청서를 제출했어. \\
        (I have already submitted my application for leave of absence.)
    \end{tabular}
    \\\cline{2-3}

    & \begin{tabular} {@{}c@{}}
        휴가 \\
        (Vacation)
    \end{tabular}
    & \begin{tabular} {@{}l@{}}
        부다페스트는 야경이 정말 아름다웠어. \\
        (Budapest was really beautiful at night.)
    \end{tabular}
    \\\cline{2-3}

    & \begin{tabular} {@{}c@{}}
        문화/예술 \\
        (Art/Culture)
    \end{tabular}
    & \begin{tabular} {@{}l@{}}
        내가 원하는 날짜에 뮤지컬 공연 예약 예매했어. \\
        (I made a reservation for a musical performance on the date I want.)
    \end{tabular}
    \\\cline{2-3}

    & \begin{tabular} {@{}c@{}}
        교통수단 \\
        (Transportation)
    \end{tabular}
    & \begin{tabular} {@{}l@{}}
        나는 자가용을 사용하는 게 편해. \\
        (I'm comfortable using my own car.)
    \end{tabular}
    \\\cline{2-3}

    & \begin{tabular} {@{}c@{}}
        사회/시사 \\
        (Social Issues)
    \end{tabular}
    & \begin{tabular} {@{}l@{}}
        어 그래서 우리나라도 지금 제주도로 난민이 꽤 들어오고 있어. \\
        (Uh, that's why there are quite a few refugees coming to Jeju Island.)
    \end{tabular}
    \\\cline{2-3}

    & \begin{tabular} {@{}c@{}}
        관혼상제 \\
        (Ceremonial Occasions)
    \end{tabular}
    & \begin{tabular} {@{}l@{}}
        내일 아는 분 결혼식에 가는데 어떻게 입고 가는 게 좋을까요? \\
        (What should I wear tommorrow for the wedding of my aquaintance?)
    \end{tabular}  
    \\\hline
    \hline
    \multirow{4}{*}[-2em]{
        \begin{tabular} {@{}c@{}}
            Guideline \\
            (for spontaneous \\
            speech)
        \end{tabular}
    }
    & 주제 (Topic) & \multicolumn{1}{c|}{스포츠/레저 (Sports/leisure)} \\ 
    \cline{2-3}
    & \begin{tabular} {@{}c@{}}
        발화 주제 \\
        (Topic sentence)
    \end{tabular}
    & 
    \begin{tabular} {@{}l@{}}
        이스포츠의 정식 스포츠 등록에 대해 어떻게 생각하시나요? \\
        (What's your opinion about adoption of E-Sports as an official event 
        in the Asian Games?)
    \end{tabular} 
    \\\cline{2-3}
    
    & \begin{tabular} {@{}c@{}}
        관련 키워드 \\
        (Keywords)
    \end{tabular}
    & \begin{tabular} {@{}l@{}}
        이스포츠, 스포츠, 아시안게임 \\
        (E-sports, Sports, Asian Games)
    \end{tabular}
    \\\cline{2-3}
    
    & \begin{tabular} {@{}c@{}}
        관련 뉴스 \\
        (Related News)
    \end{tabular}
    & \begin{tabular} {@{}l@{}}
        e스포츠는 왜 아직 스포츠가 되지 못했을까 \\
        (Why hasn't eSports become a sport yet?)
    \end{tabular} 
    \\
 \hline
\end{tabular}
}
\\
}
\end{table*}